\documentclass[12pt,preprint]{aastex}

\shorttitle{An FX software correlator for VLBI}
\shortauthors{Deller, Tingay, West, \& Bailes}

\begin{document}

\title{DiFX: A software correlator for very long baseline interferometry using multi-processor computing environments}

\author{A.T. Deller\altaffilmark{1}, S.J. Tingay, M. Bailes, \& C. West\altaffilmark{2}}
\affil{Centre for Astrophysics and Supercomputing, Swinburne University of Technology, Mail H39, P.O. Box 218, Hawthorn, Victoria 3198, Australia}

\altaffiltext{1}{co-supervised through the Australia Telescope National Facility, P.O. Box 76, Epping, NSW 1710, Australia}

\altaffiltext{2}{Current address: University of Massachusetts Amherst, Department of Astronomy, 710 North Pleasant St, Amherst, MA 01003-9305, USA}

\begin{abstract}
We describe the development of an FX style correlator for Very Long
Baseline Interferometry (VLBI), implemented in software and intended
to run in multi-processor computing environments, such as large
clusters of commodity machines (Beowulf clusters) or computers
specifically designed for high performance computing, such as
multi-processor shared-memory machines.  We outline the scientific and
practical benefits for VLBI correlation, these
chiefly being due to the inherent flexibility of software and the fact
that the highly parallel and scalable nature of the correlation task
is well suited to a multi-processor computing environment.  We suggest
scientific applications where such an approach to VLBI correlation is
most suited and will give the best returns.  We report detailed
results from the Distributed FX (DiFX) software correlator,
running on the Swinburne supercomputer (a Beowulf cluster of $\sim$300
commodity processors), including measures of the performance of the
system.  For example, to correlate all Stokes products for a 10 antenna array,
with an aggregate bandwidth of 64 MHz per station and using typical time and frequency resolution 
presently requires of
order 100 desktop-class compute nodes.  Due to the effect of Moore's
Law on commodity computing performance, the total number and cost of compute nodes 
required to meet a given correlation task continues to decrease rapidly with time.
We show detailed comparisons between DiFX
and two existing hardware-based correlators: the Australian Long Baseline Array (LBA) S2 correlator, and the NRAO Very Long Baseline Array (VLBA) correlator.  In both cases, excellent agreement was 
found between the correlators. Finally, we describe plans for the 
future operation of DiFX on the Swinburne supercomputer,
for both astrophysical and geodetic science.
\end{abstract}


\section{Introduction}
The technique of Very Long Baseline Interferometry (VLBI), as a means
to study the very high angular resolution structure of celestial radio
sources, was developed in the 1960s \citep{cla67,mor67}.  Some accounts of the early
developments in VLBI, the scientific motivations for the developments,
and technical overviews are given in \citet{fin00}. 

VLBI, as with all interferometry at radio wavelengths, hinges on the
abilty to obtain a digital representation of the electric field
variations at a number of spatially separated locations (radio
telescopes), accurately time-tagged and tied to a frequency standard.
The digitised data are transported to a single location
for processing (a correlator) and are coherently combined in order to derive
information about the high angular resolution structure of the target
sources of radio emission.  The instantaneous angular resolution $R$  
of a VLBI array in arcseconds is given by $R = 2.52\times10^{5} \frac{\lambda}{D}$, where $\lambda$
is wavelength of the radiation being observed
(typically centimetres) and $D$ is the maximum projected baseline
(the distance between radio telescopes in the array projected onto a plane perpendicular
to the source; typically thousands of kilometers).  This yields typical angular resolutions 
of order milliarcseconds.

Traditionally, the ``baseband" data (filtered, down-converted, sampled, and quantised electric field strength measurements: \citealt{tho94}) generated at each radio telescope have been recorded to
magnetic tape media, for example: the Mark I system \citep{bar67}; the Mark II system \citep{cla73}; the Mark III system \citep{rog83}, the Mark IV system \citep{whi93}; and the S2 system \citep{wie96}.  After observation, the tapes from each telescope were shipped to a purpose-built and
dedicated digital signal processor, the correlator.  A
correlator aligns the recorded data streams, corrects for various
geometrical and instrumental effects, and coherently combines the data from the different
independent pairs of radio telescopes.  The correlator output streams, known as the visibilities, are
related to the sky brightness distribution of the radio source essentially via a
Fourier transform relation \citep{tho94}.

The two fundamental operations required to combine or correlate the recorded signals are a Fourier transform (F) and a cross-multiplication (X).  The order of these operations can be interchanged to obtain the same result, leading to the so-called XF and FX correlator architectures.  A number of well-known descriptions of the theory and practise of
radio interferometry describe the technique in varying degrees of detail and elaborate upon the differences between XF and FX correlators \citep{tho94,rom99}, and the reader is referred to these texts for the details.

Both XF and FX style correlators have traditionally been highly application-specific devices, based on purpose-built integrated circuits.  In the last 20 years, Field Programmable Gate Arrays (FPGAs) have become popular in correlator designs, with one prominent example being the Very Long Baseline Array (VLBA) correlator \citep{nap94}.  FPGAs are reconfigurable or reprogrammable devices that offer more flexibility than application-specific integrated circuits (ASICs) while still being highly efficient.

This paper deals with a departure from the traditional approach of
tape-based data recording and correlation on a purpose-built processor (based on either ASICs or FPGAs).  We have developed a correlator that is based on software known as DiFX (Distributed FX), which runs within a generic multi-processor computing environment.  Such a correlator interfaces naturally to modern hard-disk data recording systems, such as the MkV system \citep{whi02} and the K5 system \citep{kon03}, that have now largely replaced tape-based recording systems.  Specifically, we have developed this software correlator to support a new disk-based VLBI recording system that has been deployed across the Australian Long Baseline Array\footnote{http://www.atnf.csiro.au/vlbi} (LBA) for VLBI.  We refer the reader to a detailed discussion of the LBA hard-disk recording system (LBADR) that appears elsewhere \citep[in preparation]{phi06}.  As our software correlator is more broadly applicable than to just the LBA, we will not dwell on the details of the LBA recording system in this paper, but rather concentrate on the characteristics, benefits, and performance of our software correlator, giving brief details of the recording system when required.  The correlator source code, binaries and instructions for use are available for download from \verb+http://astronomy.swin.edu.au/~adeller/software/difx/+.

The very first VLBI observations were in fact correlated using software on a mainframe computer.  Software correlators were developed simultaneously on an IBM 360/50 at the National Radio Astronomy Observatory (NRAO) \citep{bar67} and on an IBM 360/92 at the Goddard Space Flight Centre \citep{mor67}.  As the early experiments
quickly increased in complexity the recorded data volume also
increased and it became necessary to design custom hardware for VLBI correlation.  Recent
examples of such correlators include: the NRAO Very Long Baseline Array correlator \citep{nap94}; the Joint Institute for VLBI in Europe (JIVE) correlator \citep{cas99}; the Canadian NRC S2 correlator \citep{car99}; the Japanese VLBI Space Observatory Programme (VSOP) correlator \citep{hor00}; and the Australia Telescope National Facility (ATNF) S2 correlator \citep{wil96}. Table \ref{tab:corrcomp} compares some of the basic properties of some currently-operational hardware VLBI correlators.

Recently, the pace of development of commodity computing equipment (processors,
storage, networking etc) has outstripped increases in VLBI computational
requirements to the point that the correlation of VLBI data using relatively
inexpensive supercomputer facilities is feasible.  The correlation algorithm is ``embarrassingly parallel" and very 
well suited to such parallel computing architectures.
These facilities are not purpose-built for correlation but are inherently
multi-purpose machines, suited to a wide range of computational
problems.  

This approach to correlation gives rise to significant scientific
benefits, under certain circumstances.  The benefits stem from the basic 
characteristics of correlation, software engineering considerations, and the
computing environments.  Software is more flexible and easier to redesign
than application-specific hardware or even FPGA-based processors (although 
the programming tools for FPGAs are developing rapidly).  The highly 
parallel nature of the correlation problem, coupled with the availability
of high-level programming languages and optimised vector libraries means that a
reasonably general software correlator code can be written quickly and
be used in a variety of different computing environments with minimum
modification, or in a dynamic environment where computing resources
and/or significant scientific requirements can change rapidly with time.

However, the trade-off
for flexibility and the convenience of high-level programming tools is reduced 
efficiency for any given task, compared to an application-specific or FPGA-based solution.  
Put simply, the Non-Recoverable Engineering (NRE) costs for a software
correlator are much lower than for a hardware correlator, but the cost
per unit processing power is higher.  Thus, the limited computation needed by a small size correlator
means a software approach will be cheaper overall, while the tremendous computational
requirements of correlators on the scale required for the Expanded Very Large Array 
(EVLA) or Atacama Large Millimetre Array (ALMA) dictate that the substantial 
amounts of NRE spent optimising hardware are worthwhile, at least in 2006.

Software also has an advantage over hardware if the additional support required
for unusual or stringent VLBI experiments is impossible or impractical to implement
in an existing hardware correlator.  An example of this is given in \S \ref{subsec:scintillation}.  
Use of a software correlator in these cases, even at possibly reduced efficiency, is preferable to
the expense of building or altering dedicated hardware.

A good example of the flexibility of software correlation and its
trade-off with efficiency is spectral resolution capability.   A generic modern
CPU is capable of calculating multi-million point one-dimensional 
Fast Fourier Transforms (FFTs), allowing an FX style
software correlator utilising this CPU as a processing element to give
extremely high frequency resolution: a million spectral points across the
frequency bandwidth of an observation.

Such a correlation would be computationally intensive, as conventional CPUs are not optimised for such operations.  However, it could be carried out using exactly the same software and hardware as is used for a generic continuum experiment.  Comparison to Table \ref{tab:corrcomp} shows that such high spectral resolution is currently impossible on existing hardware correlators.  A number of limitations on particular hardware correlator implementations, such as minimum integration times, maximum input data rates, and maximum output data rates, can be overcome in a similar fashion with software correlators.

The flexibility, inexpensive nature, and ease of production of software correlators
makes them particularly useful for small to medium sized VLBI arrays, since
development times are short, costs are low, and the capabilities are
high, providing niche roles for even small facilities.  These factors
have led to a resurgence in software correlator applications in a
number of groups around the world.  In addition to the efforts
described here at the Swinburne University of Technology, a group have
developed a software correlator, mainly for geodetic VLBI, at the Communications Research Laboratory (CRL)
in Japan \citep{kon03}.  This CRL code is also used for real-time
fringe checks during observations on the European VLBI Network (EVN),
operated from JIVE\footnote{Details about the process and results can be found at http://www.evlbi.org/evlbi/tevlb8/tevlb8.html}.  Also at JIVE, 
a software correlator has been developed and used to process VLBI observations that
tracked the Huygens probe as it entered the atmosphere of Titan \citep{pog03}.  Spacecraft tracking with VLBI and
software correlation is likely to become a more recognised technique
following the Huygens success, for example for the Chinese Chang'E lunar mission\footnote{http://en.cast.cn}.  
Finally, the most ambitious example of
a software correlator is the Low Frequency Array (LOFAR) correlator, which is implemented on
an IBM BlueGene/L supercomputer containing 12,000
processors\footnote{http://www.lofar.org}.  This software correlator rivals the most
powerful hardware correlators currently operating or in the design
stage, but differs from the software correlator described in this paper in that hardware
specific optimisations and large amounts of NRE were utilised.  

The approach we used in the development of the software correlator was largely inspired by the previous success of a group at Swinburne who developed baseband signal processing software for multi-processor environments, for the purposes of pulsar studies \citep{bai03}.  A prototype software correlator developed at Swinburne is described in \cite{wes04}, with initial results described in \cite{hor06}.

In this paper we concentrate on a description of the DiFX software
correlator for VLBI developed at the Swinburne University of
Technology, motivated by the factors discussed above.  This correlator
has been used as part of the Australia Telescope National Facility (ATNF)
VLBI operations since 2005 and has now
replaced the previously used ATNF S2 correlator.  The particular
architecture we have adopted (\S 2.1, 2.2 and 2.3), is discussed only briefly, as the correlation algorithm has been discussed at length in the literature.  
\S 3 describes the DiFX correlator, including the details of the software
implementation, verification results from comparisons with two  established hardware
correlators, and performance figures-of-merit.
We illustrate some examples of specific scientific applications
that can benefit from software correlation in \S 4.  Finally, our conclusions are presented in \S 5.

\section{The FX software correlator architecture}
Many previous works develop in detail the theory of radio interferometry \citep{tho94,tho99}.  The reader is referred to these texts for a complete discussion of the technique.  Here we discuss the main steps used to implement the correlator architecture (FX) that we have adopted.

A more extensive overview of correlator operations is given in \citet{rom99}.  We do not describe the operations at the telescopes that convert the incident electric field at sky frequency to the filtered, down-converted, sampled, and digitised data streams that are recorded to disk (baseband data in our terminology).

A number of the initial operations are made on the telescope-based data streams.  A number of the later operations are baseline-based.  These two sets of operations are briefly described separately and in sequence. 

\subsection{Antenna-based operations}
\subsubsection{Alignment of telescope data streams}
\label{subsec:alignment}
To correlate data from a number of different telescopes, the changing delays between those telescopes must be calculated and used to align the recorded data streams at a predetermined point in space (in this case the geocentre) throughout the experiment. 

The Swinburne software correlator uses CALC 9\footnote{http://gemini.gsfc.nasa.gov/solve} to generate a
geometric delay model ($\tau(t)$) for each telescope in a given observation, at regular intervals (usually 1 second).  CALC models many geometric effects, including precession, nutation, ocean and atmospheric loading, and is used by many VLBI correlators including the VLBA and JIVE correlators.  These delays are
then interpolated (using a quadratic approximation) to produce accurate delays ($\Delta\tau < 1 \times 10^{-15}$ sec, compared to an exact CALC value) in double precision for any time during the course of the observation.   The estimated station clock offets and rates are added to the CALC-generated geometric delays.

The baseband data for each telescope are loaded into large buffers in memory, and the
interpolated delay model is used to calculate the accurate delay between each telescope and the centre of the Earth at any given time during the experiment.  This delay, rounded to the nearest sample, is the integer sample delay.  The difference between the delay and the integer sample delay is recorded as the antenna based fractional sample delay (up to $\pm$ 0.5 sample).  Note that the alignment of any two data streams (as opposed to a data stream alignment with the geocentre) is good to $\pm$ 1 sample.

The integer-sample delay is used to offset the data
pointer in memory and select the data to be correlated (some number of samples which is a power of 2, starting from the time of alignment).  The fractional sample error is retained to correct the phase as a function of frequency following alignment to within one sample, fringe rotation, and channelisation (\S 2.1.3).

Once the baseband data for each telescope have been selected, they are transferred to a processing node and unpacked from the
coarsely quantised representation (usually a 2-bit representation) to a floating point (single precision) representation.  From this
point on, all operations in the correlator are performed using floating point arithmetic,
in single precision unless otherwise specified.  Note that the data volume is expanded by a factor of 16 at this point.  The choice of single precision floats (roughly double the precision necessary) was dictated by the capabilities of modern CPUs, which process floats efficiently.  Using sufficient precision also avoids the small decorrelation losses incurred by optimised, low precision operations often used in hardware correlators.  This is a good example of the sacrifice of efficiency for simplicity and accuracy with a software correlator.

At this point all data streams from all telescopes are aligned to within $\pm$ 1 sample of each other and the fractional sample errors for each of the telescope data streams are recorded for later use.  A set number of samples from each telescope data stream have been selected and are awaiting processing on a common processing node (e.g. a PC in a Beowulf cluster).

\subsubsection{Fringe rotation}
Fringe rotation compensates for the changing phase difference introduced by
delaying the signal from each telescope to the geocentre after it has been downconverted to
baseband frequencies.  If the changing delay, $\tau(t)$, could be compensated for at sky frequency, fringe rotation would not be required.  This, however, is impractical.

The necessary fringe rotation function can be
calculated at any point in time by taking the sine and cosine of the
geocentric delay multiplied by the sky frequency $\nu_{0}$; it is applied via
a complex multiplication for each telescope's data stream.

Since the baseband data have already
been unpacked to a floating point representation by this stage, a floating point fringe
rotation is applied which yields no fringe rotation losses, compared, for example, to
a 6.25\% loss of signal to noise for three level digital fringe 
rotation in a two level complex correlator \citep{rob97}.

Implemented as such, fringe rotation represents a mixing operation and will result in a phase difference term which is quasi-stationary at zero phase (the desired term) and a phase sum term which has a phase rate of twice the fringe rotation function, $\sim4\pi\nu_{0}\tau(t)$.  The sum term vector averages to a (normally) negligible contribution to the correlator; for typical VLBI fringe rates (100s of kHz) and integration times (seconds) the relative magnitude of the unwanted contribution to each visibility point is $<10^{-5}$.  In a software correlator it would be simple to control the integration time so that the rapidly varying phase term is integrated over exactly an integral number of terms of phase, thus making no contribution to the correlator output.  This feature is not currently implemented in DiFX.

We have thus far described fringe rotation as a phase shift for each sample in the time domain.  If performed in this manner, we refer to the fringe rotation
as ``pre--F'' (under an FX architecture), as it has been applied before the transformation to
the frequency domain in the channelisation process (\S \ref{subsec:channel}).  In this case, the geometric delay for each
sample is interpolated using the delay model as described in \S \ref{subsec:alignment} above.

In cases where the fringe rotation to be applied changes
little from the first sample in the FFT window to the last, a minimal
amount of decorrelation is introduced by applying a single fringe
rotation for the entire window.  The decorrelation can be estimated by
$\rm sinc \left(\Delta\phi/2 \right)$, where $\Delta\phi = 2 \pi \nu_{0} \Delta\tau$ is the change in baseline phase due to
Earth rotation over the FFT window.

In this way, fringe rotation can be applied after
channelisation, which saves considerable computational effort (``post--F" fringe rotation).  For
this approach to be viable, the fringe rates should be low (ie low
frequencies and/or short baselines) and the number of channels should
be small (implying that the time range of the samples to be correlated is short compared to the fringe period).  Table \ref{tab:decorr} shows
the degree of decorrelation which would be incurred by utilising post--F
fringe rotation for a range of VLBI observation modes. This decorrelation is simple to calculate and
could be used to correct the visibility amplitudes and alter visibility weights, 
although this is not presently implemented in DiFX.  It is important to note that the use of 
post--F fringe rotation is not recommended for all situations shown in Table \ref{tab:decorr},
and indeed is only intended for use when the resultant decorrelation is $\ll 1\%$.

Post--F fringe rotation is desirable in situations where the fringe
rate is extremely low, when the double-frequency term introduced
by the mixing operation of pre--F fringe rotation is not effectively averaged to zero over the
course of an integration and makes a significant and undesirable contribution to the correlator output.  Switching from pre--F to post--F
fringe rotation would be beneficial for periods of time in
most experiments when the source traverses periods of low phase rate.  Sources near a celestial pole can have very low fringe rates for long periods of time.  Alternatively, if very short correlator integration times are used, the sum term may not integrate to zero when using pre--F fringe rotation.  Post-F fringe rotation would therefore be a natural choice in these circumstances.

It should be noted that it is possible to undertake the exact equivalent to pre--F fringe rotation in the frequency domain.  However, this would involve the Fourier transform of the fringe rotation function and a convolution in the frequency domain, which is at least as computationally intensive as the complex multiplication of the data and fringe rotation in the time domain.

DiFX implements pre--F or post--F fringe rotation as a user controlled option.

\subsubsection{Channelisation and fractional sample error correction}
\label{subsec:channel}
Once the data are aligned and phase corrected after fringe rotation, the time series data are converted into frequency series data (channelised), prior to cross multiplication.

Channelisation of the data can be accomplished using an FFT (Fast Fourier Transform) or a
digital filterbank.  If used, the filterbank is implemented in a
polyphase fashion, which essentially inserts a decomposed filter
before an FFT \citep{bel74}.  This allows the channel response to be changed from the
$\rm sinc^2$ response natural to a FX correlator to any desired function.
In practise, an approximation to a rectangle is applied, although the
length of the filter (and hence the accuracy of the approximation) is
tunable.

If pre-F fringe rotation has been applied, the data are already in
complex form, and so a complex-to-complex FFT is used.  The positive
or negative frequencies are selected in the case of upper or lower
sideband data respectively.  If post-F fringe rotation is to be
applied, the data are still real and so a more efficient
real-to-complex FFT may be used.  This is possible due to the
conjugate symmetry property of an FFT of a real data series.  In this
case, lower sideband data may be recovered by reversing and
conjugating the resultant channels.

The final station-based operation is fractional-sample correction \citep{rom99}.
This step is considerably easier in an FX correlator than an XF
implementation, since the conversion to the frequency domain before
correlation allows the fractional error to be corrected exactly, assuming the error to be constant over an FFT length.  
This is equivalent to the assumption made for post-F fringe rotation, but is considerably less stringent since the
phase change is proportional to the subband bandwidth, rather than sky frequency as in the case of fringe rotation.
The frequency domain correction manifests itself as a slope in the phase as a function of frequency across the observed bandwidth.

Thus, after channelisation, a further complex multiplication is
applied to the channels, correcting the fractional sample error.  In
the case of post-F fringe rotation, the fringe rotation value is added
to the fractional-sample correction and the two steps are performed
together.

Either simple FFT or digital polyphase filter bank channelisation can be selected as a user controlled option in DiFX.

\subsection{Baseline-based operations}
\subsubsection{Cross multiplication of telescope data streams}
For each baseline, the channelised data from the telescope pair are
cross-multiplied on a channel by channel basis (after forming the complex conjugate for the channelised data from one telescope) to yield the frequency domain complex visibilities that are
the fundamental observables of an interferometer.  This is repeated
for each common band/polarisation on a baseline, and for all baselines.  If dual
polarisations have been recorded for any given band, the cross-polarisation
terms can also be multiplied, allowing polarisation information for the target source to be recovered.

\subsubsection{Integration of correlated output}
Once the above cycle of operations has been completed, it is repeated
and the resulting visibilities accumulated (complex added) until a set accumulation
time has been reached.  The number of "good" cycles per telescope is recorded, which 
could form the basis of a data weighting scheme, although weights are not 
currently recorded in DiFX.  Generally, on each cycle the input time
increment is equal to the corresponding FFT length (twice the number of spectral points), but it is also possible to overlap FFTs.  This
allows more measurements of higher lags and greater sensitivity to
spectral line observations, at the cost of increased computation.  In
this way, the limiting time accuracy with which accumulation can be
performed is equal to the FFT length divided by the overlap factor.  A
caveat to this statement is discussed in \S 3.4.

\subsubsection{Calibration for nominal telescope $T_{sys}$}
Cross multiplication, accumulation and normalisation by the antenna autocorrelation spectra gives the complex cross power spectrum for each baseline, representing the correlated fraction of the geometric mean of the powers detected at each telescope.  To obtain the correlated power in units of Jy, the cross power spectra (amplitude components) should be scaled by the geometric mean of the powers received at each telescope measured in Jy i.e. the $T_{sys}$ in Jy routinely measured at each antenna.  Calibration based on the measured $T_{sys}$ is typically performed as a post-correlation step in AIPS\footnote{http://www.aoc.nrao.edu/aips} or a similar data analysis package, and so a nominal value for the $T_{sys}$ for each telescope is applied at the correlator.  In addition, a scaling factor to compensate for decorrelation due to the coarse quantisation of the baseband data is applied.  This corrects the visibility amplitudes, but of course cannot recover the lost signal to noise. For the 2-bit data typically processed, this scaling factor is 1/0.88 in the low-correlation limit \citep{coo70}.  The relationship becomes non-linear at high correlation and the scaling factor approaches unity as the correlation coeffient approaches unity.  The correction for high-correlation cases can be applied in post-processing, generally at the same time as the application of measured $T_{sys}$ values.

\subsubsection{Export of visibility data}
Once an accumulation interval has been reached, the visibilities must
be stored in a useful format.  Presently, the software
correlator supports RPFITS\footnote{http://www.atnf.csiro.au/computing/software/rpfits.html} as the output format.  RPFITS files can
be loaded into analysis packages such as AIPS, CASA\footnote{http://casa.nrao.edu/}, or MIRIAD\footnote{http://www.atnf.csiro.au/computing/software/miriad} for data
reduction.  Ancillary information is included in the RPFITS file along with the complex visibilities, time stamps, and (u,v,w) coordinates.  The RPFITS standard supports the appending of a 
data weight to each spectral point, but DiFX does not currently record weights.
In the future, it is planned to add additional widely used output formats, such as FITS-IDI\footnote{http://www.aoc.nrao.edu/aips/FITS-IDI.html}.

\subsection{Special processing operations: pulsar binning}
Pulsed signals are dispersed as they travel through the interstellar
medium (ISM), resulting in a smearing of the pulse arrival time in
frequency.  In order to
correct for the dispersive effects of the ISM, DiFX
employs incoherent dedispersion \citep{vou02}.  This allows the
visibilities generated by the correlator to be divided into pulse
phase bins.  Unlike hardware correlators which typically allow only a
single on/off bin, or else employ $2^N$ bins of fixed width, DiFX
allows an arbitrary number of bins placed at
arbitary phase intervals.  The individual bins can be written out separately in
the RPFITS file format to enable investigation of pulse phase dependent effects,
or can be filtered within the correlator based on a priori pulse profile information.

To calculate which phase bin a visibility at a given frequency and
time corresponds to, the software correlator requires information on
the pulsar's ephemeris, which is supplied in the form of one or more
``polyco" files containing a polynomial description of apparent pulse
phase as a function of time.  These are generated using the pulsar
analysis program TEMPO\footnote{http://pulsar.princeton.edu/tempo/reference\_manual.html}, and require prior timing of a pulsar.  Additional software has been written by the authors to
verify the pulsar timing, using the generated polyco files and the baseband data (in MkV, LBA or K5 format) from an experiment, allowing phase bins to be accurately set before correlation.

For VLBI observations of pulsars, it is usually desirable
to maximise the signal to noise of the observations by binning the
visibilities based on the pulse phase, and applying a filter to the
binned output based on the signal strength in that phase.  Typically
this filter is implemented as a binary on/off for each phase bin.  Using the
pulse profile generated from the baseband data of an observation,
however, DiFX allows a user-specified number of
bins to be generated and a filter applied based on pulse strength $\times$ bin
width, allowing the maximum theoretical retrieval of signal, as described below.  This
also reduces the output data volume, since only an ``integrated
on-pulse'' visibility is retained, rather than potentially many phase
bins.

Consider observing a single pulse, divided into $M$ equally spaced phase bins.
Let the pulsar signal strength as a function of phase bin
be $S(m)$, and the noise in single phase bin to be $Z \times \sqrt{M}$, where $Z$ is the baseline sensitivity
for an integration time of a single pulse period.  When all bins are summed (effectively no binning), the S/N ratio will be:

\begin{equation}
\frac{\sum_{m=0}^{M}S(m)}{Z}
\end{equation}

\noindent
as the signal adds coherently while the noise adds in quadrature.
For a simple on/off gate accepting only bins $m_{1}$ to $m_{2}$, the S/N ratio will be:

\begin{equation}
\frac{\sum_{m=m_{1}}^{m_{2}}S(m)}{\sqrt{\sum_{m=m_{1}}^{m_{2}}\left( Z\times\sqrt(M)\right)^{2} }}
\end{equation}

Finally, for the case where each bin is weighted by the pulse signal strength in that bin,
the S/N ratio will be:

\begin{equation}
\frac{\sum_{m=0}^{M}\left(S(m)\right)^{2}}{\sqrt{\sum_{m=0}^{M}\left( S(m) \times Z \times \sqrt(M)\right)^{2} }}
\end{equation}

For a Gaussian shaped pulse, this allows a modest improvement in recovered signal to 
noise of 6\% compared to an optimally placed single on/off bin.
On a more complicated profile, such as a Gaussian main pulse with a Gaussian interpulse at half the amplitude, the improvement in recovered signal to noise increases to 21\%.

\section{Software correlation on the Swinburne Beowulf cluster - a case study}
\subsection{The cluster computing environment}
The Swinburne University of Technology supercomputer is a $\sim$300 processor Beowulf
cluster, that is a mixture of commodity off-the-shelf desktop and server style PCs, 
connected via a gigabit ethernet network.  In particular, the supercomputer
has five sub-clusters, each with 48 machines.  Four sub-clusters are made up of single processor 3.2 GHz, Pentium 4 PCs with 1 GB of RAM per machine, while one sub-cluster is made up of dual processor Xeon servers, each with 2 GB of RAM per machine.  The cluster is continuously upgraded and fully replaced approximately every 3--4 years.  The software correlation code must operate in this multi-user, multi-tasking, and highly dynamic environment.

\subsection{Structure of the DiFX code}
\label{sec:codeoutline}
DiFX is written in C$++$, but makes heavy
use of the optimised vector processing routines provided by the Intel
Performance Primitive (IPP) library\footnote{http://www.intel.com/cd/software/products/asmo-na/eng/perflib/ipp/index.htm}.  The use of this optimised vector library results in a factor of several 
performance gain on the Intel CPUs, compared to non-optimised vector code. Data transfer is
handled via the Message Passing Interface (MPI) standard\footnote{http://www-unix.mcs.anl.gov/mpi/}.  The mpich implementation of MPI is used\footnote{http://www-unix.mcs.anl.gov/mpi/mpich1/}.

\begin{figure}
\includegraphics[angle=0, scale=1.85]{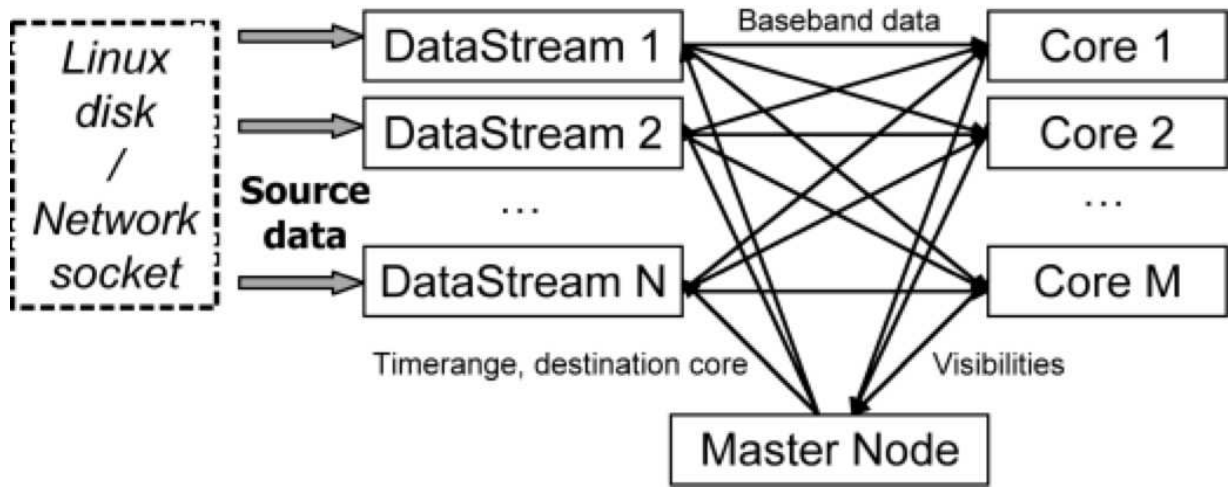}
\caption{Overview of the software correlator architecture.  Data is loaded into memory from a disk or
network connection by Datastream nodes.  These nodes are directed by a Master node to send data from 
given time ranges (typically several ms) to the processing elements (Core nodes).  The processed data
are sent to the master node for long-term accumulation and storage on disk.}
\label{fig:corrlayout}
\end{figure}

Figure \ref{fig:corrlayout} shows the high-level class structure of DiFX,
along with the data flow. The correlation is managed by a master node (FxManager),
which instructs data management nodes (Datastream) to send time ranges
of baseband data to processing nodes (Core).  The data are then processed
by the Core nodes, and the results sent back to the FxManager.
Double buffered, non-blocking communication is used to avoid latency delays
and maximise throughtput.  Both the Datastream and Core classes can be (and have
been) extended to allow maximum code re-use when handling different
data formats and processing algorithms.  The Core nodes make use of an allocatable number
of threads to maximise performance on a heterogenous cluster.

The Datastream nodes can read the baseband data into their memory buffers from a local disk,
a network disk or a network socket.  Once the data are loaded into the datastream buffer, the
remainder of the system is unaware of its origin.  This is one of the most powerful aspects of
this correlator architecture, meaning the same correlator can easily be used for production
disk-based VLBI correlation and real-time eVLBI testing, where the data is transmitted in real
time from the telescopes to the correlator over optical fibre.  Real-time eVLBI  operational modes have been tested using DiFX, transmitting data in real-time from the three ATNF telescopes (Parkes, ATCA, and Mopra) to computing resources at the Swinburne University of Technology and the University of Western Australia in Perth (a Cray XD-1 utilising Opteron processors and on-board Xilinx FPGAs).  The software correlator then correlates the transmitted data in real-time.  A full account of the new eVLBI capabilities of the Australian VLBI array will be presented elsewhere \citep[in preparation]{phi06}.

\subsection{Operating DiFX}
DiFX is controlled via an interactive Graphical User Interface (GUI), which
calls the various component programs and helper scripts.  The primary purpose of the GUI is to
facilitate easy editing of the text files which configure the
correlator, run external programs such as the delay model generator, 
and provide feedback while a job is running.  
Two files are necessary to run the actual correlator program.  The first
is an experiment configuration file, containing tables of stations,
frequency setups, etc, analogous to a typical hardware correlator job configuration script.
The second file contains the list of compute nodes on which the
correlator program will run.

While it is possible to run all tasks required to operate
the correlator manually, in practise they are organised via the GUI.
This consists of running a series of helper applications
from the GUI to generate the necessary input for the correlator.
These include a script to extract experiment information from the VLBI
exchange (VEX) file used to configure and schedule the telescopes at observe time, a delay and (u,v,w)
generator which makes use of CALC 9, and scripts to extract the
current load of available nodes.  Pulsar--specific information such
as pulse profiles and bin settings can also be loaded.  This information is presented via
the GUI and adjustments to the configuration, such as selection computational resources to be used, can be made before launching a correlation job.

In the future it is planned to incorporate some real-time feedback of
amplitude, phase and lag information from the current correlation via the GUI.
This would be similar to the visibility spectra displays available continuously
at connected-element interferometers.

\subsection{Performance}
In order to keep every compute node used in the correlation fully loaded,
they must be kept supplied with raw data.  If this condition is satisfied, we
have a CPU-limited correlation, and the addition of further nodes will result
in a linear performance gain.  In practise, however, at some
point obtaining data from the data source (network socket or disk) and 
transmitting it across the local network to the
processing nodes will no longer occur quickly enough,
and the correlation becomes data-limited rather
than CPU-limited.  Correct selection of correlation parameters, and
good cluster design, will minimise the networking overhead imposed on
a correlation job, and ensure that all compute nodes are fully utilised.  This is discussed in \S \ref{sec:network} below,
and performance profiles for the CPU-limited case are presented in \S \ref{sec:perfnumbers}.

\subsubsection{Networking considerations}
\label{sec:network}
As described in \S \ref{sec:codeoutline}, double-buffered communications to the processing nodes
are used to ensure that nodes are never idle as long as sufficient aggregate networking capability is available.
The use of MPI communications adds a small but unavoidable overhead to data transfer, meaning the maximum
throughput of the system is slightly less than the maximum network capacity on the most heavily loaded data path.

There are two significant data flows: out of each Datastream and into
the FxManager.  For any high speed correlation, there will be more
Core nodes than Datastream nodes, so the aggregate rate into a Core will be lower
than that out of a Datastream.  The flow out of a Core is a factor of $N_{\rm cores}$ times
lower than that into the FxManager node.

If processing in real time (when processing time equals observation time), the rate
out of each Datastream will be equal to the recording rate, which can be up to 1 Gbps with
modern VLBI arrays and is within the capabilities of modern commodity ethernet equipment.
The rate into the FxManager node will be equal to the product of the recording rate, the compression ratio, and the
number of Cores, where the compression ratio is the ratio of data into
a Core to data out of a Core.  This is determined by the number of
antennas (since number of baselines scales with number of antennas
squared), the number of channels in the output cross-power spectrum, the number
of polarisation products correlated, and the integration time used before sending data back to
the FxManager node.

It is clearly desirable to maximise the size of data messages sent to a
core for processing, since this minimises the data rate into the
FxManager node for a given number of Cores.  However, if the messages are too large,
performance will suffer as RAM capacity is exceeded.  Network latency
may also become problematic, even with buffering.  Furthermore, it
should be apparent that in this architecture, the Cores act
as short-term accumulators (STAs), with the manager performing the
long term accumulation.  The length of the STA sets the minimum
integration time.  It is important to note, however, that the STA
interval is entirely configurable in the software correlator, to be as
short as a single FFT, although network bandwidth and latency are likely to be limiting
factors in this case.

For the majority of experiments it is possible to set a STA length
which satisfies all the network criteria and allows the Cores to be
maximally utilised.  For combinations of large numbers of
antennas and very high spectral and time resolution, however, it is
impossible to set an STA which allows a satisfactorily low return data
rate to the FxManager node.  In this case, real time processing
of the experiment is not possible without the installation of 
additional network and/or CPU capacity on the FxManager node.

It is important to emphasise that although it is possible to find
experimental configurations for which the software correlator suffers
a reduction in performance, these configurations would be impossible
on existing hardware correlators.  If communication to the FxManager node
is limiting performance, it is also possible to parallelize a disk-based experiment by
dividing an experiment into several time ranges and processing these time ranges
simultaneously, allowing an aggregate processing rate which equals
real time.  This is actually one of the most powerful aspects of the
software correlator, and one which would allow scheduling of
correlation to always ensure the cluster was being fully utilised.

\subsubsection{CPU-limited performance}
\label{sec:perfnumbers}
\begin{figure}
\includegraphics[angle=270,scale=0.65]{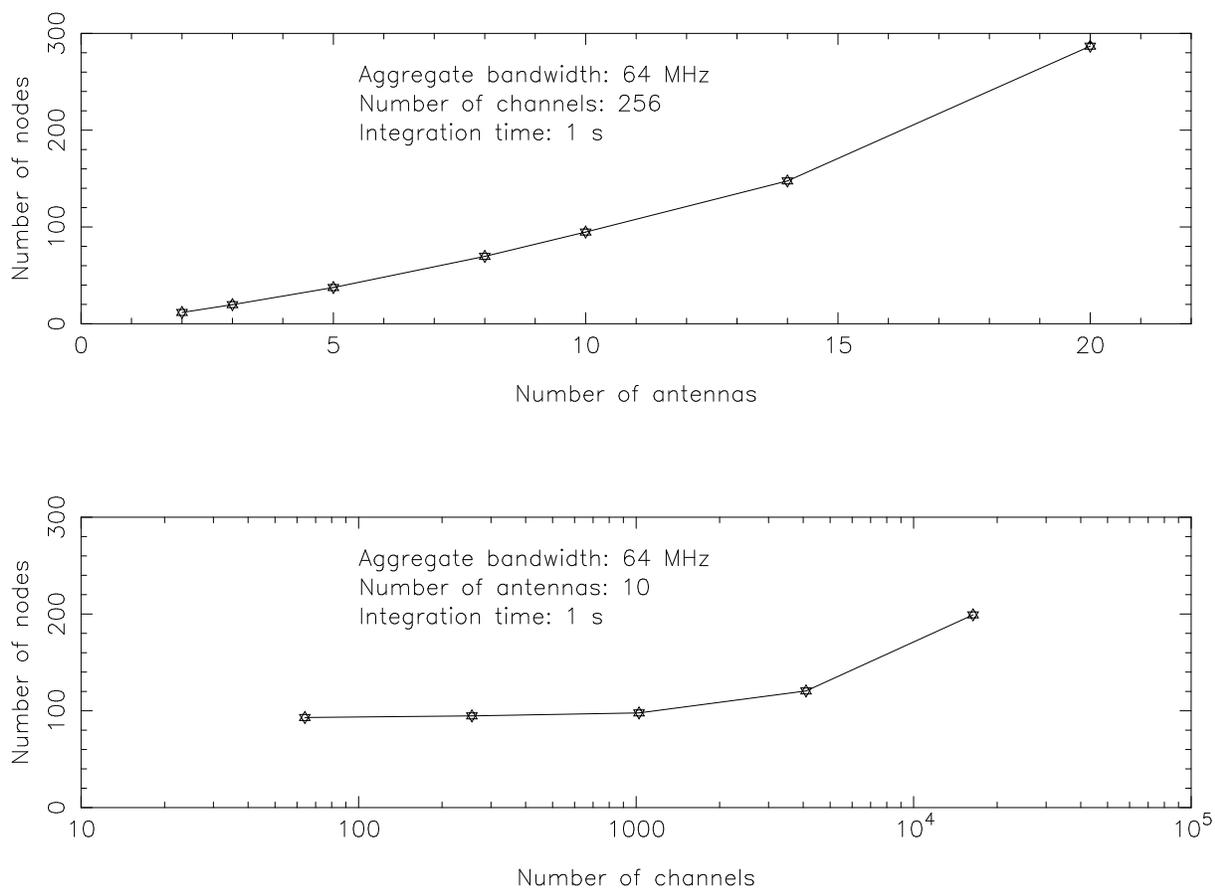}
\caption{Benchmark data showing the computational requirements of DiFX to correlate in real-time, as described in the text.  The nodes are single core 3.2 GHz Pentium processors with 1 GB RAM, and in both benchmarks 64 MHz of total bandwidth per station was correlated with a 1 second integration period.  Top panel shows the scaling of computational requirements with number of antenna, using 256 spectral points per 8 MHz subband.  Bottom panel shows the scaling of computional requirements with spectral points per subbband for a ten station array.}
\label{fig:benchmarks}
\end{figure}

Figure \ref{fig:benchmarks} shows the results of performance testing on the Swinburne
cluster (using the 3.2 GHz Pentium 4 machines and the gigabit ethernet network) for different
array sizes and spectral resolutions.  The results shown in Figure \ref{fig:benchmarks} were obtained for data for which the aggregate bandwidth was 64 MHz, broken up into 8 bands each of 8 MHz bandwidth (4 $\times$ dual polarisation 8 MHz bands: data were 2-bit sampled: antenna data rate 256 Mbps).  Node requirements for real-time operation are extrapolated from the compute time on an 8 node cluster.  The correlation integration time is 1 second and all correlations provide all four polarisation products.
RAM requirements per node ranged from 10 -- 50 MB depending on spectral resolution, showing
that large amounts of RAM are unnecessary for typical correlations.
It can be seen that even a modestly sized commodity cluster can 
process a VLBI-sized array in real time at currently available data rates.

\subsection{Correlator comparison results}

\subsubsection{Comparison with ATNF S2 correlator}
\label{subsec:s2comp}
Observations to provide data for a correlator comparison between the Swinburne software correlator and the ATNF S2 correlator were undertaken on March 12, 2006, with the following subset of the LBA: Parkes (64 m), ATCA (phased array of 5 $\times$ 22 m), Mopra (22 m), Hobart (26 m).

Data from these observations were recorded simultaneously to S2 tapes and the LBADR disks \citep[in preparation]{phi06} during a 20 minute period, UT 02:30--02:50, corresponding to a scan on a bright quasar (PKS 0208$-$512).  The data recorded corresponded to two 16 MHz bands, right circular polarisation (RCP), in the frequency ranges 2252 $-$ 2268 MHz and 2268 $-$ 2284 MHz.

The data recorded on S2 tapes were shipped to the ATNF LBA S2 correlator \citep{rob97} at ATNF headquarters and processed.  The data recorded to LBADR disks were shipped to the Swinburne University of Technology supercomputer and processed using the software correlator.

At both correlators identical $T_{sys}$ values in Jy were specified for each antenna and applied in order to produce nominally calibrated visibility amplitudes.  Further, both correlators used identical clock models, in the form of a single clock offset and linear rate as a function of time per antenna.  Finally, the data were processed at each correlator using 2 second correlator integration times and 32 spectral channels across each 16 MHz band.

Different implementations of the CALC-based delay generation were used at each correlator, meaning small differences exist in the delay models used, leading to differences in the correlated visibility phase.  We have calculated the delay model differences and subtracted the phase due to differential delay model in the following discussion.

\begin{figure}
\includegraphics[angle=270,scale=0.65]{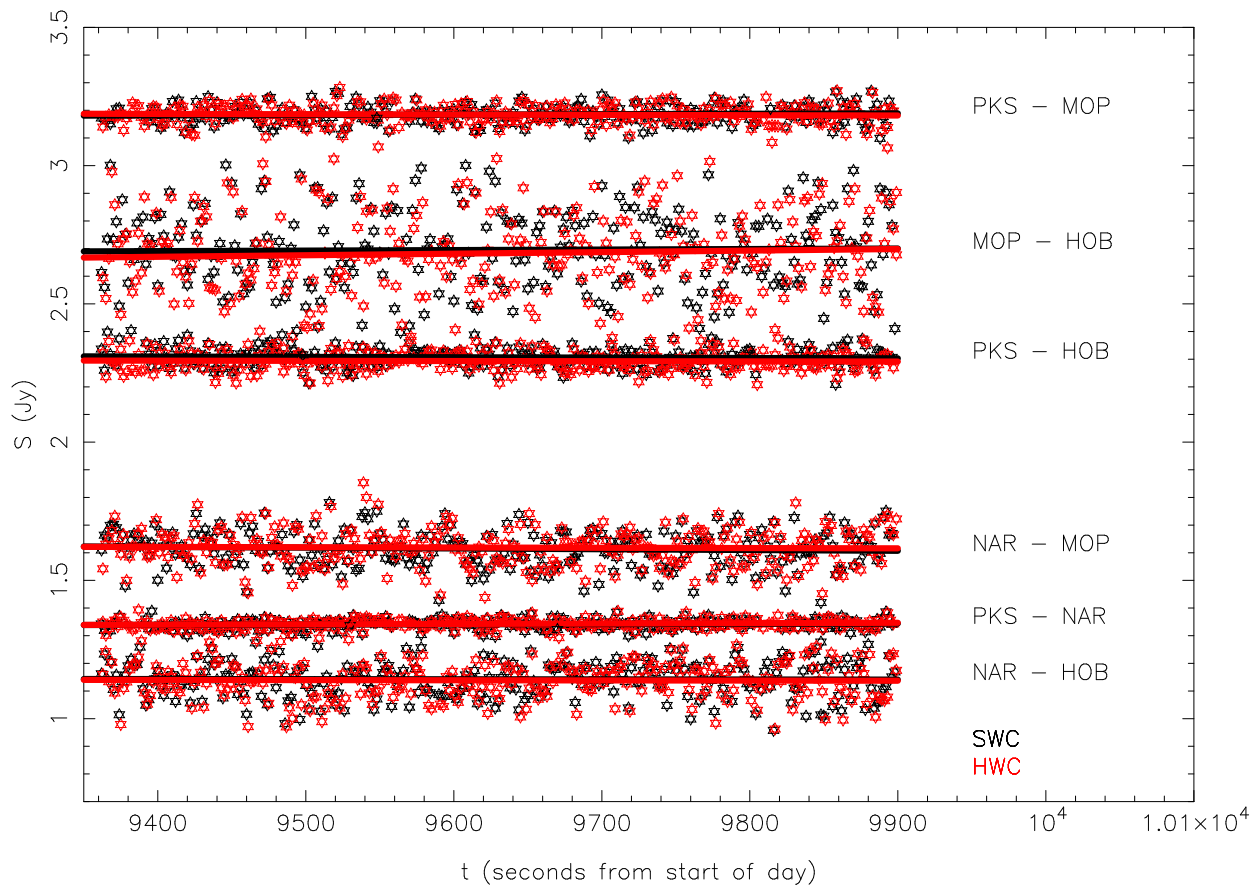}
\caption{S2 (red) and DiFX (black) visibility amplitude vs time for the 2252 -- 2268 MHz band on the source PKS 0208$-$512, as described in the text (PKS $=$ Parkes; MOP $=$ Mopra; HOB $=$ Hobart; NAR $=$ ATCA).  Symbols represent the actual visibilities produced by the correlators, while the lines represent linear least-squares fits to the visibilities (one line per dataset).}
\label{fig:s2amptime}
\end{figure}

\begin{figure}
\includegraphics[angle=270,scale=0.65]{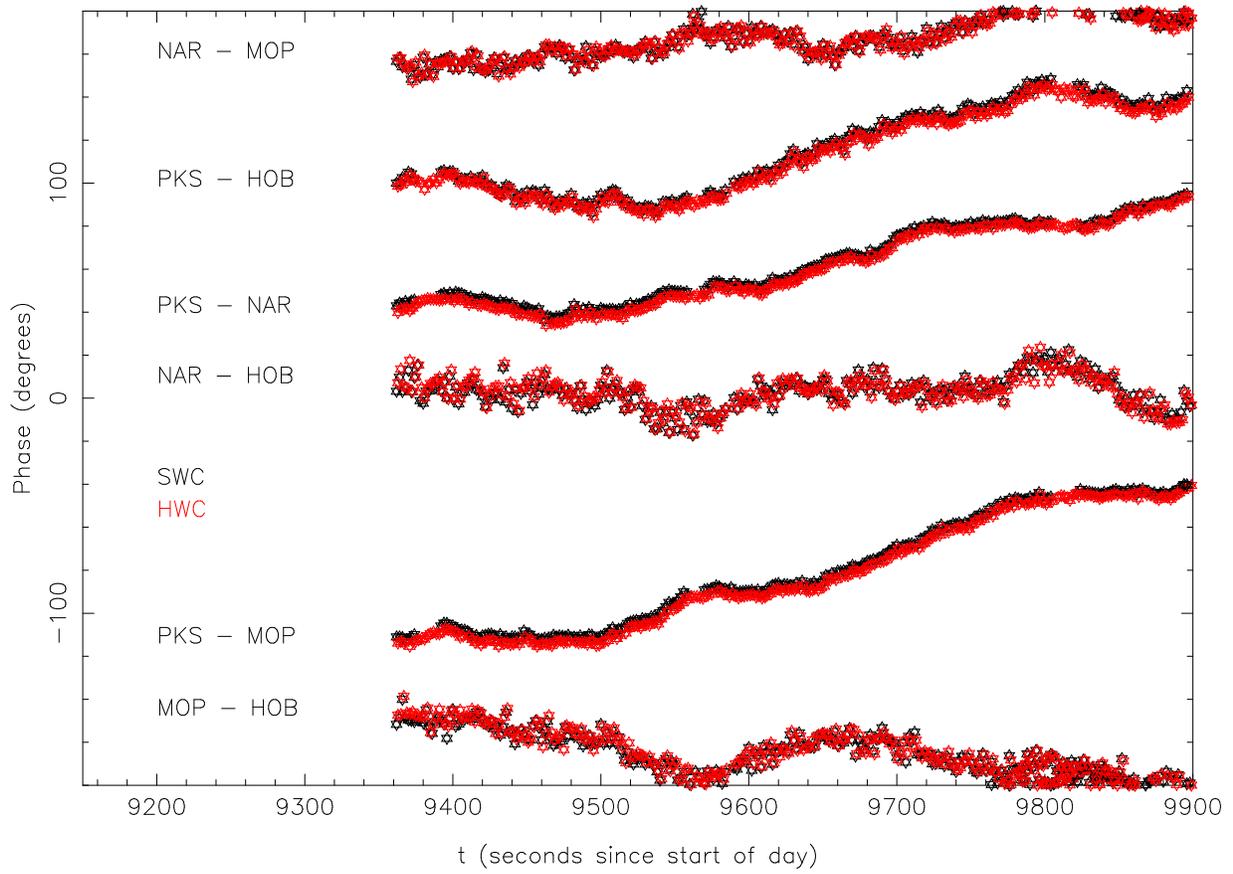}
\caption{S2 (red) and DiFX (black) visibility phase vs time for the 2252 -- 2268 MHz band on the source PKS 0208$-$512, as described in the text.  Antenna labels as in Figure 2 above.  The PKS-NAR baseline has been shifted by $-50\deg$ for clarity.}
\label{fig:s2phasetime}
\end{figure}

\begin{figure}
\includegraphics[angle=270,scale=0.65]{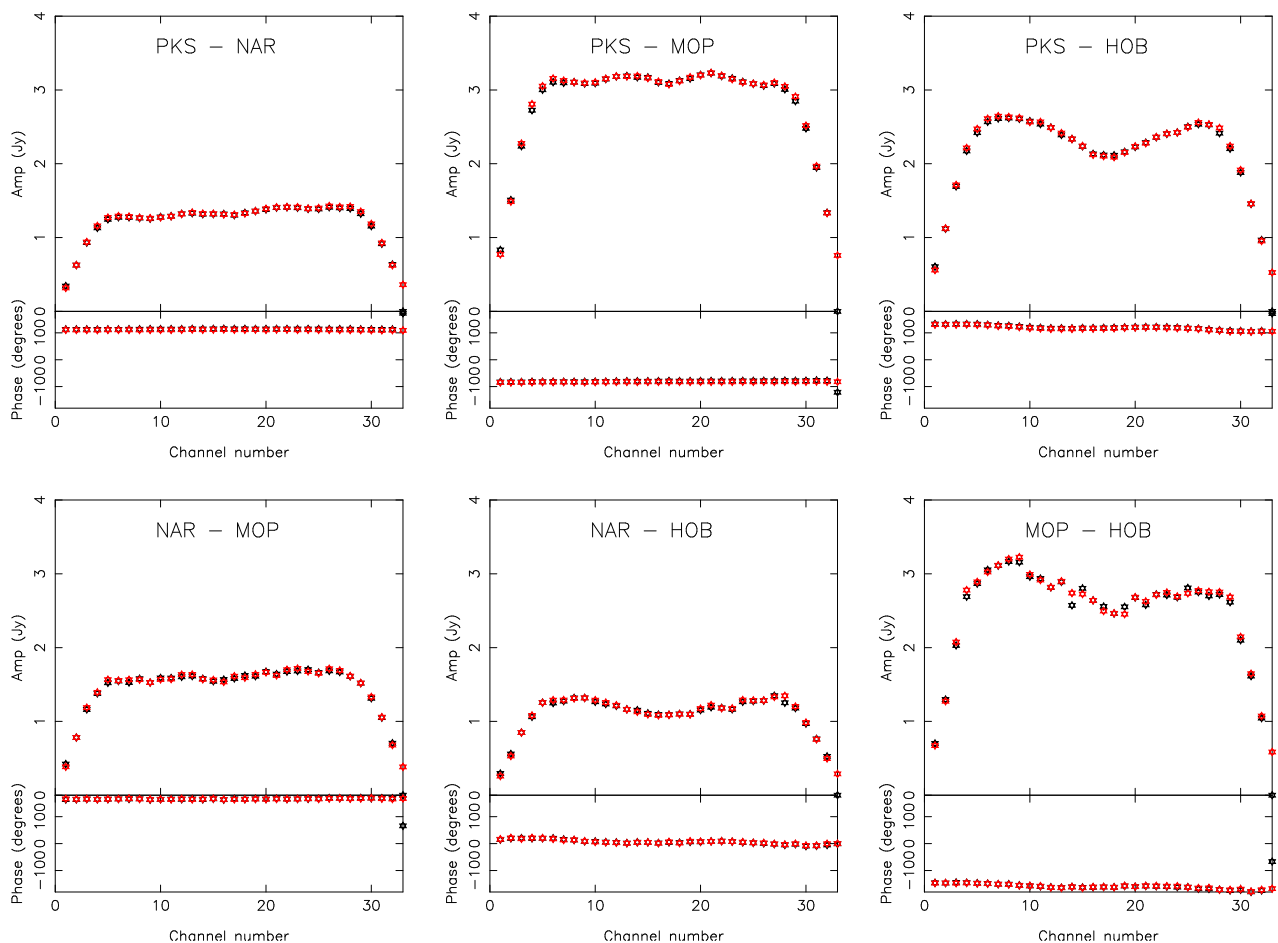}
\caption{S2 (red) and DiFX (black) visibility amplitude and phase vs frequency data for the 2252 -- 2268 MHz band on the source PKS 0208$-$512, as described in the text.  Antenna labels as in Figure 2 above.  The S2 data has been corrected for fractional-sample error decorrelation at the band edges as described in the text.}
\label{fig:s2ampfreq}
\end{figure}

From both correlators, RPFITS format data were output and loaded into the MIRIAD software \citep{sau} for inspection and analysis.  The data from the two correlators are compared in a series of Figures below (Figures \ref{fig:s2amptime} -- \ref{fig:s2ampfreq}).

Figure \ref{fig:s2amptime} shows the visibility amplitudes for all baselines from both correlators as a function of time, over the period 02:36:00 - 02:45:00 UT, for one of the 16 MHz bands (2252 $-$ 2268 MHz).  These amplitudes represent the vector averaged data over the frequency channel range 10 $-$ 21 (to avoid the edges of the band).  The data for each baseline were fit to a first order polynomial model ($S(t)=\frac{dS}{dt}t + S_{0}$, where $S$ is the flux density in Jy, $t$ is the offset in seconds from UT 02:40:30, and $S_{0}$ is the extrapolated flux density at time UT 02:40:30, using a standard linear least squares routine.  The root mean square (RMS) variation around the best fit model was calculated for each baseline.  The fitted models are shown in Figure \ref{fig:s2amptime} and show no significant differences between the S2 correlator and the software correlator.  Further, the calculated RMS for each baseline agrees very well between DiFX and the S2 correlator, as summarised in 
Table \ref{tab:s2comp}.

Figure \ref{fig:s2phasetime} shows the visibility phase as a function of time for each of the six baselines in the array.  Again the data represent the vector averaged correlator output over the frequency channel range 10 $-$ 21 within the 2252 $-$ 2268 MHz band.  As discussed above, small differences between the delay models used at each correlator have been taken into account as part of this comparison.  

Figure \ref{fig:s2ampfreq} shows a comparison of the visibility amplitudes and phases as a function of frequency in the 2252 $-$ 2268 MHz band.  The data represented here result from a vector average of the two datasets over a two minute time range, UT 02:40:00 $-$ 02:42:00.  Since the S2 correlator is an XF -- style correlator, it cannot exactly correct fractional sample error in the same manner as an FX correlator such as DiFX, as the channelisation is performed after accumulation.  The coarse (post-accumulation) fractional sample correction leads to decorrelation at all points except the band center, up to a maximum of $\sim10\%$ at the band edges on long baselines where the geometric delay changes by a sample or more over an integration period.  We have corrected for this band edge decorrelation in the S2 correlator amplitudes in Figure \ref{fig:s2ampfreq}.

\subsubsection{Comparison with the VLBA correlator}

Data obtained as part of a regular series of VLBA test observations were used as a basis for a correlator comparison between the software correlator and the VLBA correlator \citep{nap94}.  The observations were made on 2006 August 05 using the Brewster, Los Alamos, Mauna Kea, Owens Valley, Pie Town, and Saint Croix VLBA stations.  One bit digitised data sampled at the Nyquist rate for four dual polarisation bands, each of 8 MHz bandwidth, were recorded using the Mk5 system \citep{whi03}.  The four bands were at centre frequencies of 2279.49, 2287.49, 2295.49, and 2303.49 MHz.  The experiment code for the observations was MT628 and the source observed was 0923$+$392, a strong and compact active galactic nucleus.  Approximately two minutes of data recorded in this way was used for the comparison.

The Mk5 data were correlated on the VLBA correlator and exported to FITS format files.  The data were also shipped to the Swinburne supercomputer and correlated using the software correlator, the correlated data exported to RPFITS format files.  In both cases, no scaling of the correlated visibility amplitudes by the system temperatures were made at the correlators.  The visibilities remained in the form of correlation coefficients for the purposes of the comparison i.e. a system temperature of unity was used to scale the amplitudes.  Each 8 MHz band was correlated with 64 spectral points, and an integration time of 2.048 seconds was used.

The VLBA correlator data were read into AIPS using FITLD with the parameter DIGICOR$=$1.  The DIGICOR parameter is used to apply certain scalings to the visibility amplitudes for data from the VLBA correlator.  Further, to obtain the most accurate scaling of the visibility amplitudes, the task ACCOR was used to correct for imperfect sampler thresholds, deriving corrections to the antenna-based amplitudes of $\sim0.5\%$.  These ACCOR corrections were applied to the data and the data were written to disk in FITS format.

The software correlator data were read directly into AIPS and then written to disk in the same FITS format as the VLBA correlator data.  No corrections to amplitude or phase of the software correlated data were made in AIPS.

The VLBA correlator data and the software correlator data were both imported into MIRIAD for inspection and analysis, using the same software as used for the comparison with the LBA correlator described above.  RCP from the 2283.49 -- 2291.49 MHz band over the time range UT 17:49:00 $-$ 17:51:00 was used in all comparison plots below.

Since the delay models used by the VLBA and software correlators differ at the picosecond level, as is the case for the comparison with the LBA data in \S \ref{subsec:s2comp}, differences in the visibility phase exist between the correlated datasets.  As with the LBA comparison, we have compensated for the phase error due to the delay models differences in the following comparison.

\begin{figure}
\includegraphics[angle=270,scale=0.65]{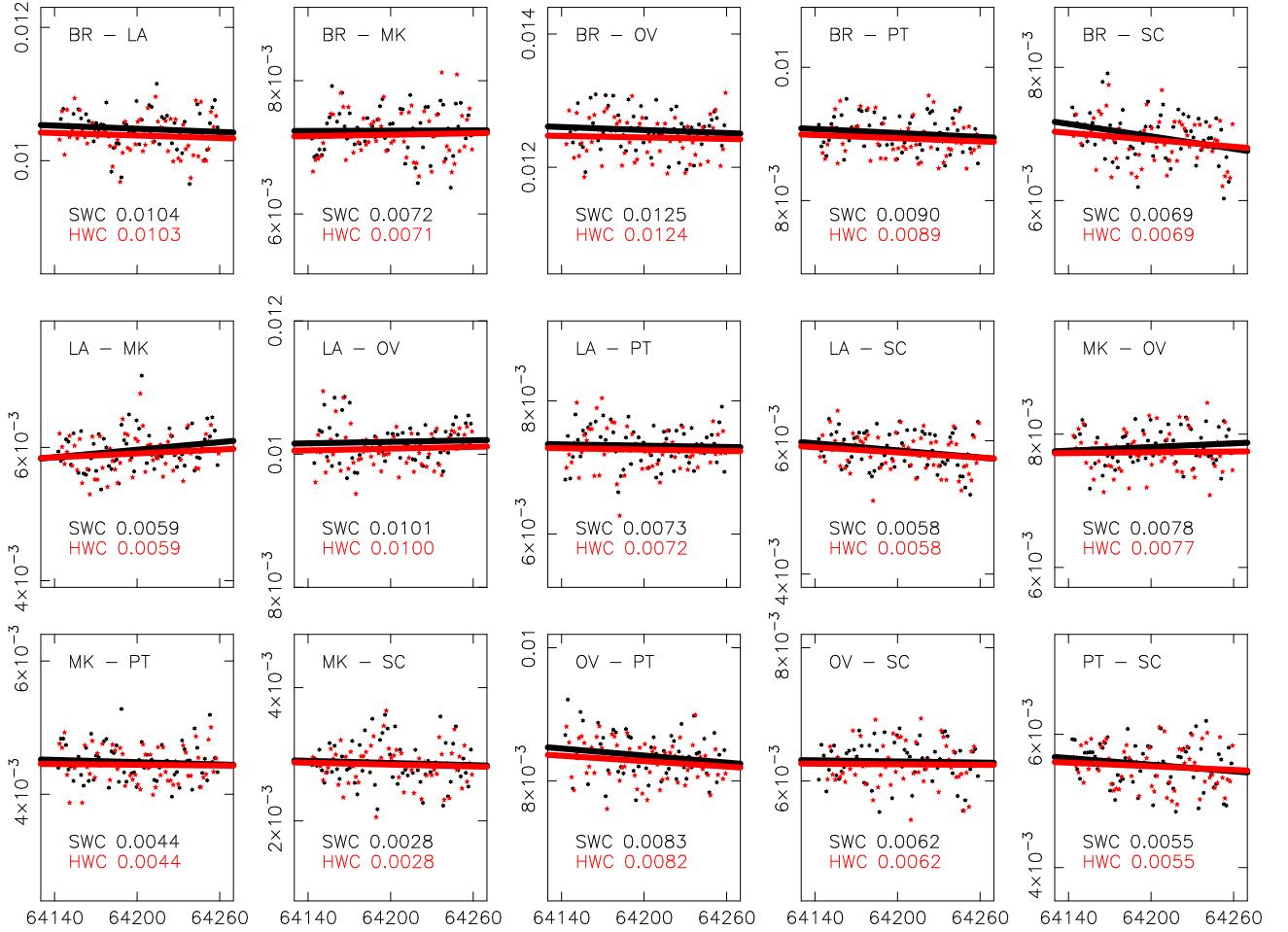}
\caption{VLBA correlator (red) and DiFX (black) visibility amplitude vs time for the 2283.49 -- 2291.49 RCP band from the VLBA test observation MT628, as described in the text.  The units of time are seconds from UT 00:00:00, and the amplitude scale is correlation coefficient.  Symbols represent the actual visibilities produced by the correlators, while the lines represent linear least-squares fits to the visibilities.  The text annotation on each panel lists the average correlation coefficient amplitude for each correlator over the time period, as tabulated in Table \ref{tab:vlbacomp}.}
\label{fig:vlbaamptime}
\end{figure}

Figure \ref{fig:vlbaamptime} shows the visibility amplitudes for all baselines from both correlators as a function of time.  These amplitudes represent the vector averaged data over the frequency channel range 10 $-$ 55 (to avoid the edges of the band).  The data for each baseline were fit to a first order polynomial model ($S(t)=\frac{dS}{dt}t + S_{0}$, where $S$ is the correlation coefficient, $t$ is the offset in seconds from UT 17:50:00, and $S_{0}$ is the extrapolated correlation coefficient at time UT 17:50:00) using a standard linear least squares routine.  The root mean square (RMS) variation around the best fit model was calculated for each baseline.  The fitted models are shown in Figure \ref{fig:vlbaamptime} and show no significant differences between the VLBA correlator and the software correlator.  Further, the calculated RMS for each baseline agrees very well between the VLBA correlator and the software correlator.  The results of the comparison are summarised in Table \ref{tab:vlbacomp}.

\begin{figure}
\includegraphics[angle=270,scale=0.65]{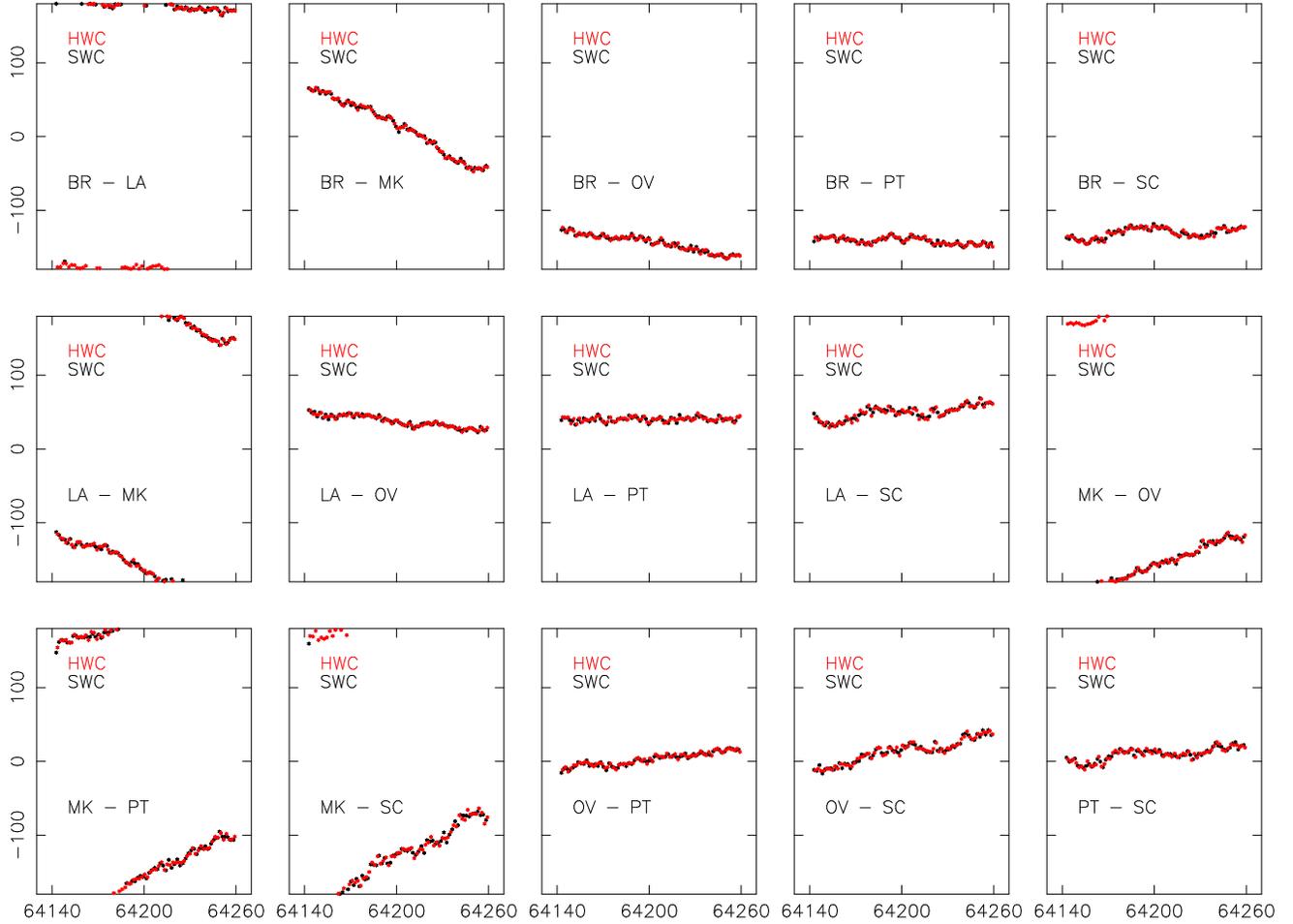}
\caption{VLBA correlator (red) and DiFX (black) visibility phase vs time for the 2283.49 -- 2291.49 RCP band from the VLBA test observation MT628, as described in the text.  The units of time are seconds from UT 00:00:00, and phase is displayed in degrees.}
\label{fig:vlbaphasetime}
\end{figure}

Figure \ref{fig:vlbaphasetime} shows the visibility phase as a function of time for each of the fifteen baselines in the array.  Again the data represent the vector averaged correlator output over the frequency channel range 10 $-$ 55 within the band.  As discussed above, small differences between the delay models used at each correlator cause phase offsets between the two correlators, and have been taken into account as part of this comparison. 

\begin{figure}
\includegraphics[angle=270,scale=0.65]{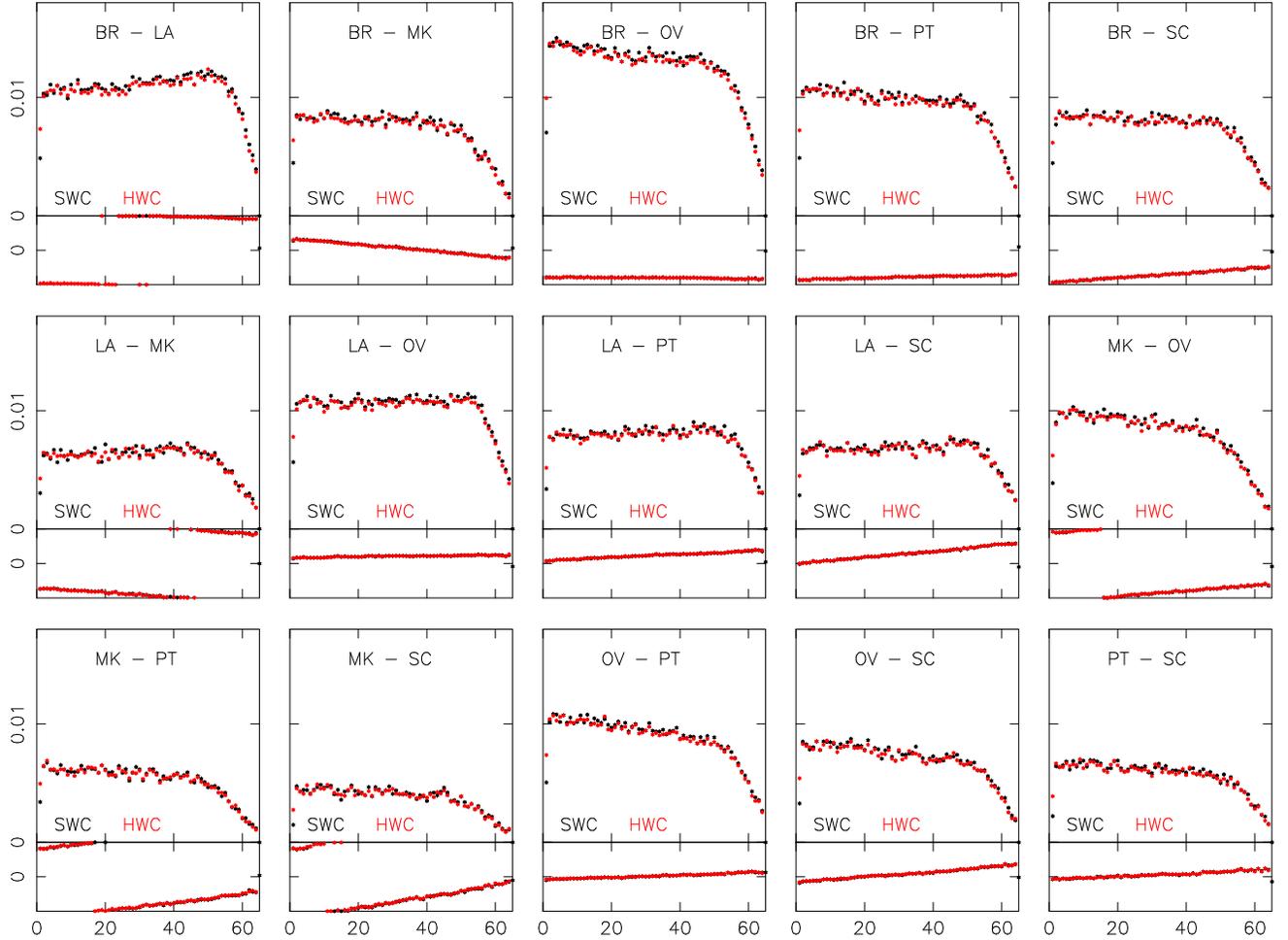}
\caption{VLBA correlator (red) and DiFX (black) visibility amplitude and phase as a function of frequency for the 2283.49 -- 2291.49 RCP band from the VLBA test observation MT628, as described in the text.  The vertical scale for correlation coefficient amplitude on each panel is 0 -- 0.018, while the phase scale spans $\pm 180\deg$.  The horizontal scale for each panel displays channels 0--64.}
\label{fig:vlbaampfreq}
\end{figure}

Figure \ref{fig:vlbaampfreq} shows a comparison of the visibility amplitudes and phases as a function of frequency in the band.  The data represented here result from a vector average of the two datasets over a two minute time range.  Figures \ref{fig:vlbaamptime}, \ref{fig:vlbaphasetime} and \ref{fig:vlbaampfreq} show that the results obtained by the VLBA correlator and DiFX agree to within the RMS errors of the visibilities in each case, as expected.

\section{Scientific applications of the Swinburne software correlator}

\subsection{High frequency resolution spectral line VLBI}
As mentioned in the introduction, an attractive feature of software
correlation is the ease with which very high spectral resolution
correlation can be undertaken.  This is particularly useful for
studies of spectral line sources such as masers when mapping the distribution of the masing regions and their kinematics i.e. near black holes in galactic nuclei \citep{gre95}.  

\begin{figure}
\includegraphics[angle=270,scale=1.4]{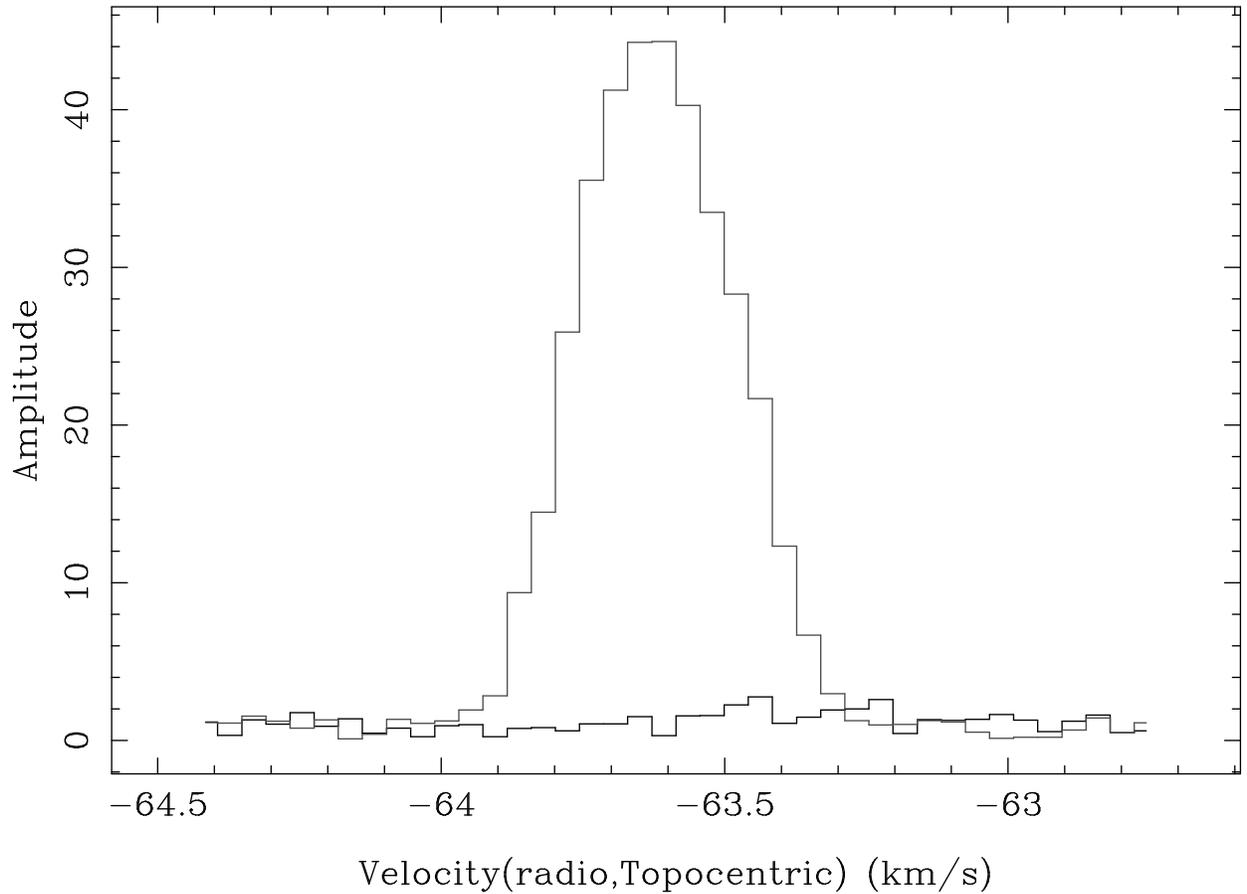}
\caption{A two minute average of the ATCA -- Parkes cross-power spectrum taken from the software correlated data for the OH maser G345$-$0.2, as described in the text.  The velocity resolution is 0.038 km/s at the central frequency of 1.72 GHz.  The light gray line showing strong maser emission represents the LCP data and the dark gray line with little emission represents the RCP data.  The maser is highly circularly polarised.}
\label{fig:maser}
\end{figure}

Figure \ref{fig:maser} shows a spectrum obtained from an LBA observation of the OH maser G345$-$0.2.  These observations were made with an array consisting of the ATCA (phased array of 5 $\times$ 22 m), Parkes (64 m), and Mopra (22 m), recording data from a dual-polarised (RCP and LCP) 4 MHz band onto hard disk.  The data were correlated using the software correlator with 16,384 frequency channels across the 4 MHz band, corresponding to 0.25 kHz per channel or 0.038 km/s velocity resolution at 1.72 GHz.

These results compare with recent very high spectral resolution work done with the VLBA.  Fish et al. (2006) observed OH masers with the VLBA, using a 62.5 kHz bandwidth and 512 channels across this band to obtain channel widths of 0.122 kHz or 0.02 km/s velocity resolution.  The velocity resolution of this correlated dataset is almost twice as good as that shown in Figure \ref{fig:maser}.  However, the VLBA bandwidth is only 0.016 times the bandwidth of the observations shown in Figure \ref{fig:maser}.

If required, DiFX could have correlated these data with 32,768 channels, 65,536 channels or even higher numbers of channels.  As mentioned in the introduction, the only penalty is compute time on a resource with a fixed number of processing elements.  DiFX therefore has a clear advantage over existing hardware correlators in terms of producing very high spectral resolution over wide bandwidths.  This capability is useful if the velocity distribution of an ensemble of masers in a field is broad and cannot be contained in a single narrow bandwidth.

\subsection{Correlation for wide fields of view}
An application that takes advantage of the frequency and time
resolution of the software correlator output is wide field imaging.
To image a wide field of view, avoiding the effects of time and
bandwidth smearing, high spectral and temporal resolution is required
in the correlator visibility output.  For example, at VLBI resolution
(40 mas), to image the full primary beam of an Australia Telescope Compact Array (ATCA)
antenna (22 m diameter) at a frequency of 1.4 GHz, requires a time
resolution of the correlator output of 50 ms and a frequency
resolution of 4 kHz (allowing a 0.75 \% smearing loss at the FWHM of the primary beam).

Neither the JIVE nor the VLBA hardware correlators can achieve such
high frequency or time resolution for continuum experiments, but DiFX can be
configured for such modes in an identical manner to a normal continuum
experiment.

\subsection{Pulsar studies}
\label{subsec:scintillation}
As compact sources with high velocities, pulsars make excellent
testbeds with which to probe the structure of the interstellar medium (ISM).  Scintillation
due to structure in a scattering screen between the observer and the
pulsar causes variations in the interferometric visibilities, which have some
dependence on time and frequency (e.g. \citealp{hew85}).  Naturally, pulsar binning is
advantageous in these studies for maximising signal to noise ratios.

The most stringent requirement for useful studies of pulsar
scintillation, however, is that of extremely high frequency
resolution.  \citet[in preparation]{bri06} have recently demonstrated the capabilities of DiFX for this type of analysis with observations of the pulsar B0834$-$04.  The NRAO Green Bank Telescope (100 m), Westerbork (14 $\times$ 25 m), Jodrell Bank (76 m), and Arecibo (305 m) were used to provide an ultra-sensitive array at 327 MHz.  The data were recorded using the Mk5 system and correlated on the Swinburne software correlator.  The main requirement on the correlation was 0.25 kHz wide frequency channels,
over the broadest bandwidth available, to maximise signal to noise.  For these observations a 32 MHz band was available.  The Swinburne software correlator therefore correlated the data with 131,072 frequency channels across the band.

No existing hardware correlator can provide such a high frequency resolution over such a wide bandwidth.  Full details of the interpretation of the B0834$-$04 software correlated data will be available in \citet[in preparation]{bri06}.  Shown in Figure \ref{fig:scintillation} is a section of the dynamic spectrum from this observation which shows the scintillation structure as functions of time and frequency.

\subsection{Geodetic VLBI}
In addition to astronomical VLBI, the software correlator can also be deployed for geodetic VLBI.  Compared to astronomical VLBI, geodetic VLBI has additional requirements, including different output formats and the frequent use of sub-arraying.  The flexibility and capabilities of the software correlator are well-matched to this task.

The software correlator has been tested on geodetic datasets obtained using the Mk5 recording system, consisting of 16 frequency bands.  These tests form the basis of a geodetic correlation comparison between the software correlator and the geodetic correlator of the Max Plank Institut of Radioastronomie in Bonn, Germany.  Full results of this correlator comparison will be reported elsewhere \citep[in preparation]{tin06}.

In particular, in Australia a new three-station geodetic VLBI array has been funded as part of the geospatial component of the Federal Government's National Collaborative Research Infrastructure Scheme (NCRIS).  This scheme provides for three new geodetic VLBI stations of 12 m diameter, Mk5 recording systems, and a modified version of the software correlator described in this paper.  The modifications necessary to convert DiFX into a geodetic correlator consist of the addition of phase calibration tone extraction, a streamlined interface to scan-by-scan correlation for sub-arraying, and a capability to produce visibilities in a format convenient for geodetic post-processing.

The new Australian geodetic VLBI array will participate in global geodetic observations, as well as undertaking experiments internal to the Australian tectonic plate.

\section{Conclusions}

In this paper we have outlined the main benefits of software correlation for small to medium sized VLBI arrays.  They are:

\begin{itemize}

\item The development of software correlation is rapid and does not depend on an intimate knowledge of digital signal processing hardware, just the algorithms;

\item The software is flexible and scalable to accommodate a very broad range of interferometric modes of observation, including many which cannot be supported by existing ASIC-based hardware correlators.  Software correlators are therefore ideal for novel experiments with very special requirements.  The main trade-off for improved performance with a software correlator is the increase in compute time for a fixed number of processing elements, or the addition of extra processing elements;

\item The software can easily incorporate data recorded using mixed disk-based recording hardware;

\item Medium to large multi-processor computing facilities are available at almost all university and government research institutions, allowing users easy entry into VLBI correlation;

\item The correlation algorithm is highly parallel and very well suited to a parallel multi-processor computing environment;

\item The cost of commodity computing continues to fall with time, making large parallel computing facilities more powerful and less expensive;

\item Once written, the code can be ported to a wide range of platforms and recompiled with minimal effort.

\end{itemize}

We have discussed the implementation of the DiFX software correlator on a standard Beowulf cluster at the Swinburne University of Technology and have provided performance figures-of-merit for this implementation, showing that relatively large numbers of telescopes and relatively high data rates can be correlated in ``real-time" using numbers of machines that do not exceed the capabilities of moderate to large Beowulf clusters.  Clear trade-offs are possible in many areas of performance.  For example, if real-time operation is not important it is possible to dramatically reduce the number of processing elements.

We have also showed the results of comprehensive testing of the software correlator, comparing it output to that of two established hardware correlators, the S2 correlator of the Australian Long Baseline Array, operated by  the ATNF, and the VLBA correlator.  The correlator comparisons of visibility amplitude and phase as functions of time and frequency verify that DiFX is operating correctly for astronomical VLBI observations.

DiFX now supports all Australian VLBI observations and some global VLBI experiments, at data rates up to 1 Gbps per telescope.  The DiFX code can be downloaded from \verb+http://astronomy.swin.edu.au/~adeller/software/difx/+.  A number of scientific programs have already been supported by the software correlator and are briefly discussed here.  Further, a modified version of the software correlator will be used to support a new VLBI array in Australia, dedicated to local and global geodetic observations.

\acknowledgements

\begin{figure}
\includegraphics[scale=1.75]{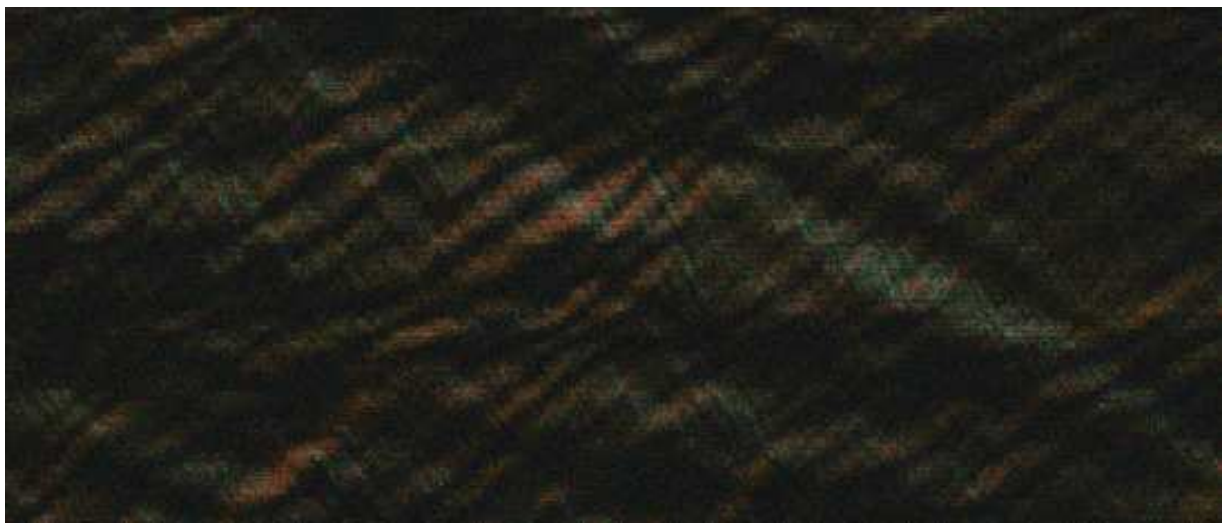}
\caption{The cross-power dynamic spectrum showing scintillation variations for the pulsar B0834$-$04 on the Green Bank Telescope -- Arecibo baseline.  Brightness represents the visibility amplitude and colour represents the visibility phase.  Increasing frequency runs left to right and increasing time runs top to bottom.  This section of the dynamic spectrum represents just 5\% of the time span and 0.5\% of the bandwidth of the observation (330 seconds and 160 kHz).}
\label{fig:scintillation}
\end{figure}

This work has been supported by the Australian Federal Government's Major National Research Facilities program, the Australian Research Council's (ARC) Strategic Research Initiatives (eResearch) scheme, and the ARC's Discovery Projects scheme.  ATD is supported via a Swinburne University of Technology Chancellor's Research Scholarship and a CSIRO postgraduate scholarship.  The Long Baseline Array is part of the Australia Telescope which is funded by the Commonwealth of Australia for operation as a National Facility managed by CSIRO.  The National Radio Astronomy Observatory is a facility of the National Science Foundation operated under cooperative agreement by Associated Universities, Inc.  We wish to thank Walter Brisken for kindly making available Figure \ref{fig:scintillation} prior to publication, the NRAO (Walter Brisken, Craig Walker, Jon Romney) for making available data for the VLBA correlator comparison and Gary Scott for correlating the LBA S2 data for the comparison with the ATNF S2 correlator.

\clearpage

\begin{deluxetable}{lccccccc}
\tabletypesize{\tiny}
\setlength{\tabcolsep}{0.025in}
\tablecaption{Comparison of existing hardware correlator parameters\label{tbl-1}}
\tablehead{
\colhead{Correlator} & \colhead{Type} & \colhead{Maximum telescopes} & \colhead{Maximum channels} & \colhead{Minimum integration time} & \colhead{Maximum input data rate }& \colhead{Maximum output data rate} & \colhead{Pulsar binning} \\
&&(in one correlator pass)&(per baseline)&\colhead{(ms)}&\colhead{(Mbps)}&\colhead{(MB/s)}&
}
\startdata
VLBA\tablenotemark{A}	&FX	&20	& 2048				&131.072				&   256	&1	&yes\\
JIVE	\tablenotemark{B}	&XF	&16	& 2048\tablenotemark{C}	&125\tablenotemark{D}	&  1024  	& 6\tablenotemark{E}&no\\
ATNF S2\tablenotemark{F}&XF	& 6	& 8192\tablenotemark{G}	&2000				&128	&0.064	& yes \\
\enddata
\tablenotetext{A}{http://www.vlba.nrao.edu/astro/obstatus/current/node28.html}
\tablenotetext{B}{http://www.jive.nl/correlator/status.html}
\tablenotetext{C}{for up to 8 telescopes}
\tablenotetext{D}{when using half the correlator}
\tablenotetext{E}{data in lag space}
\tablenotetext{F}{http://www.atnf.csiro.au/vlbi/correlator/}
\tablenotetext{G}{0.5 MHz bandwidth, 2 products}
\label{tab:corrcomp}
\end{deluxetable}

\clearpage

\begin{deluxetable}{ccccc}
\setlength{\tabcolsep}{0.03in}
\tabletypesize{\small}
\tablecaption{Maximum decorrelation incurred due to ``Post-F" fringe rotation}
\tablehead{
\colhead{Observation} & \colhead{Max. baseline} & \colhead{Frequency} & \colhead{\# channels/16MHz band} & \colhead{Max. decorrelation} \\
&\colhead{(km)}&\colhead{(MHz)}&&\colhead{(\%)}
}
\startdata
LBA low frequency continuum & 1400 & 1600 & 128 & 0.003 \\
LBA high frequency continuum & 1700 & 8400 & 128 & 0.13 \\
VLBA low frequency continuum & 8600 & 1600 & 128 & 0.12 \\
VLBA high frequency continuum & 8600 & 22200 & 128 & 21.1 \\
LBA water masers & 1700 & 22200 & 1024 & 47.6 \\
\enddata
\label{tab:decorr}
\end{deluxetable}

\clearpage

\begin{deluxetable}{ccccc}
\tablecaption{Linear fit parameters for visibility amplitude vs time for DiFX and the LBA S2 correlator, with 95\% confidence limits}
\tablehead{
\colhead{Baseline} & \colhead{Offset$_{\rm DiFX}$ (Jy)} & \colhead{Offset$_{\rm LBA}$ (Jy)}  & \colhead{Slope$_{\rm DiFX}$ ($\mu \rm Jy$ s$^{-1}$)} &\colhead{Slope$_{\rm LBA}$ ($\mu \rm Jy$ s$^{-1}$)}
}
\startdata
PKS - NAR & $1.341 \pm 0.030$ & $1.343 \pm 0.028$ & $\phantom{-}10 \pm 13\phantom{0}$ & $\phantom{-}14 \pm 12\phantom{0}$\\
PKS - MOP & $3.185 \pm 0.058$ & $3.185 \pm 0.063$ & $\phantom{-}14 \pm 24\phantom{0}$ & $-11 \pm 26\phantom{0}$\\
PKS - HOB & $2.307 \pm 0.058$ & $2.293 \pm 0.061$ & $-12 \pm 24\phantom{0}$  & $-\phantom{0}6 \pm 24\phantom{0}$\\
NAR - MOP & $1.616 \pm 0.109$ & $1.619 \pm 0.114$ & $-27 \pm 43\phantom{0}$ & $-10 \pm 45\phantom{0}$\\
NAR - HOB & $1.142 \pm 0.111$ & $1.139 \pm 0.116$ & $-\phantom{0}3 \pm 44\phantom{0}$ & $-\phantom{0}5 \pm 46\phantom{0}$\\
MOP - HOB & $2.694 \pm 0.256$ & $2.681 \pm 0.257$ & $\phantom{-}18 \pm 101$ & $\phantom{-}56 \pm 101$\\
\enddata
\label{tab:s2comp}
\end{deluxetable}

\clearpage

\begin{deluxetable}{ccccc}
\tablecaption{Linear fit parameters for visibility amplitude (in units of correlation coefficient) vs time for DiFX and the VLBA correlator, with 95\% confidence limits}
\tablehead{
\colhead{Baseline} & \colhead{Offset$_{\rm DiFX}$} & \colhead{Offset$_{\rm VLBA}$}  & \colhead{Slope$_{\rm DiFX}$ ($s^{-1} \times 10^{-6}$)} &\colhead{Slope$_{\rm VLBA}$ ($s^{-1} \times 10^{-6}$)}
}
\startdata
BR - LA & $0.0104 \pm 0.0004 $& $0.0103 \pm 0.0005$ & $-0.8 \pm 1.7$ & $-0.9 \pm 1.7$ \\
BR - MK & $0.0072  \pm 0.0005 $& $0.0071 \pm 0.0006$ & $\phantom{-}0.1 \pm 1.8$ & $\phantom{-}0.5 \pm 2.0$ \\
BR - OV & $0.0125  \pm 0.0005 $& $0.0124 \pm 0.0005$ & $-0.7 \pm 1.7$ & $-0.5 \pm 1.8$ \\
BR - PT & $0.0090  \pm 0.0004 $& $0.0089 \pm 0.0004$ & $-1.0 \pm 1.3$ & $-1.2 \pm 1.5$ \\
BR - SC & $0.0069  \pm 0.0005 $& $0.0069 \pm 0.0005$ & $-3.1 \pm 2.0$ &  $-2.5 \pm 1.8$ \\
LA - MK & $0.0059  \pm 0.0005 $& $0.0059 \pm 0.0005$ & $\phantom{-}1.9 \pm 1.7$ &   $\phantom{-}1.4 \pm 1.7$ \\
LA - OV & $0.0101  \pm 0.0005 $&  $0.0100 \pm 0.0005$ &  $\phantom{-}0.4 \pm 1.7$ & $\phantom{-}0.6 \pm 1.7$ \\
LA - PT & $0.0073  \pm 0.0005 $&  $0.0072 \pm 0.0005$ & $-0.3 \pm 1.7$ & $-0.5 \pm 1.8$ \\
LA - SC & $0.0058  \pm 0.0004 $& $0.0058 \pm 0.0004$ & $-1.8 \pm 1.5$ & $-1.9 \pm 1.5$ \\
MK - OV & $0.0078  \pm 0.0004 $&  $0.0077 \pm 0.0005$ & $\phantom{-}0.9 \pm 1.5$ & $\phantom{-}0.3 \pm 1.8$ \\
MK - PT & $0.0044  \pm 0.0004 $& $0.0044 \pm 0.0004$ & $-0.6 \pm 1.7$ & $ -0.3 \pm 1.5$ \\
MK - SC & $0.0028  \pm 0.0005 $&$ 0.0028 \pm 0.0005$ & $-0.6 \pm 1.8$ & $ -0.7 \pm 1.7$ \\
OV - PT & $0.0083  \pm 0.0005 $& $0.0082 \pm 0.0005$ & $-1.8 \pm 1.8$ & $ -1.9 \pm 1.7$ \\
OV - SC & $0.0062  \pm 0.0005 $& $0.0062 \pm 0.0005$ &$-0.3 \pm 1.8$ & $ -0.2 \pm 1.8 $\\
PT - SC & $0.0055  \pm 0.0005 $& $0.0055 \pm 0.0005$ & $-1.7 \pm 2.0$ & $ -1.3 \pm 1.8$ \\
\enddata
\label{tab:vlbacomp}
\end{deluxetable}

\clearpage

\clearpage

\clearpage

\clearpage

\clearpage

\clearpage

\clearpage

\clearpage

\clearpage

\clearpage

\begin{thebibliography}{}
\bibitem[Bailes(2003)]{bai03} Bailes, M. 2003, ASP Conf.~Ser.~302: Radio Pulsars, 302, 57
\bibitem[Bare et al.(1967)]{bar67} Bare, C. et al. 1967, Science, 157, 189
\bibitem[Bellanger \& Daguet(2004)]{bel74}  Bellanger, M. \& Daguet, J.,  IEEE Trans. Commun. Com-22(9), 1199–1205
\bibitem[Brisken et al.(2007)]{bri06} Brisken, W. et al. 2007, in preparation
\bibitem[Carlson et al.(1999)]{car99} Carlson, B.R. et al. 1999, PASP, 111, 1025
\bibitem[Casse(1999)]{cas99} Casse, J.L. 1999, New Ast. Rev., 43, 503
\bibitem[Clark(1973)]{cla73} Clark, B.G. 1973, Proc. IEEE, 61, 1242
\bibitem[Clark, Cohen \& Jauncey(1967)]{cla67} Clark, B.G., Cohen, M.H. \& Jauncey, D.L. 1967, \apj, 149, L151
\bibitem[Cooper(1970)]{coo70} Cooper, B.~F.~C.\ 1970, 
Australian Journal of Physics, 23, 521
\bibitem[Finley \& Goss(2000)]{fin00} 2000, Radio Interferometry: The Saga and the Science,'' Proceedings of a Symposium Honoring Barry Clark at 60, ed. D. G. Finley \& W. M. Goss, NRAO Workshop Number 27, Associated Universities Inc.
\bibitem[Harp(2002)]{har02} Harp, G.R. 2002, in Advanced Telescope and Instrumentation Control Software II. eds L. Hilton.  Proceedings of the SPIE, vol 4848, 1
\bibitem[Hewish et al.(1985)]{hew85} Hewish, A., Wolszczan, 
A., \& Graham, D.~A.\ 1985, \mnras, 213, 167 
\bibitem[Horiuchi et al.(2006)]{hor06} Horiuchi, S. et al. 2006, \apjs (submitted)
\bibitem[Horiuchi et al.(2000)]{hor00} Horiuchi, S. et al. 2000, Advances in Space Research, 26, 625
\bibitem[Greenhill et al.(1995)]{gre95} Greenhill, L.J. et al. 1995, \apj, 440, 619
\bibitem[Kondo et al.(2003)]{kon03} Kondo, T. et al. 2003, in New technologies in VLBI, ASP Conf. Ser., Vol. 306. ed Y.C. Minh, San Francisco, CA: Astronomical Society of the Pacific
\bibitem[Moran et al.(1967)]{mor67} Moran, J.M. et al. 1967, Science, 157, 676
\bibitem[Napier et al.(1994)]{nap94} Napier, P.J., Bagri, D.S., Clark, B.G., Rogers, A.E.E., Romney J.D., Thompson, A.R. \& Walker, R.C., Proc. IEEE, 82, 658
\bibitem[Phillips et al.(2007)]{phi06} Phillips, C. et al. 2007, in preparation
\bibitem[Pogrebenko et al.(2003)]{pog03} Pogrebenko, S. 2003, in Workshop on Planetary Probe Atmospheric Entry and descent Trajectory Analysis and Science, ed A. Wilson, ESA
\bibitem[Roberts(1997)]{rob97} Roberts, P.P. 1997, Astron. Astrophys. Suppl. Ser., 126, 379
\bibitem[Rogers et al.(1983)]{rog83} Rogers, A.E.E. et al. 1983, Science, 219, 51
\bibitem[Romney(1999)]{rom99} Romney, J.~D.\ 1999, ASP 
Conf.~Ser.~180: Synthesis Imaging in Radio Astronomy II, 180, 57
\bibitem[Sault, Teuben, \& Wright(1995)]{sau} Sault, R.J., Teuben, P.J. \& Wright, M.C.H. 1995, ASPC, 77, 433
\bibitem[Tingay et al.(2007)]{tin06} Tingay, S.J. et al 2007, in preparation
\bibitem[Thompson, Moran \& Swenson(1994)]{tho94} Thompson, A.R., Moran, J.M. \& Swenson, G.W. 1994, Interferometry and Synthesis in Radio Astronomy, Kreiger Publishing Company
\bibitem[Thompson(1999)]{tho99} Thompson, A.~R.\ 1999, ASP 
Conf.~Ser.~180: Synthesis Imaging in Radio Astronomy II, 180, 11 
\bibitem[Vo\^{u}te et al.(2002)]{vou02} Vo\^{u}te, J.L.L. et al. 2002, A\&A, 385, 733
\bibitem[West(2004)]{wes04}West, C. 2004, M.Sc. thesis, Swinburne University of Technology
\bibitem[Whitney(2003)]{whi03} Whitney, A.R. 2003, ASPC, 306, 123
\bibitem[Whitney(2002)]{whi02} Whitney, A.R. 2002, in Proceedings of the 6th European VLBI Network Symposium, eds. Ros, E., Porcas, R.W., Lobanov, A.P., \& Zensus, J.A., 41
\bibitem[Whitney(1993)]{whi93} Whitney, A.R. 1993, in IAU Symp. 156: Developments in Astrometry and their Impact on Astrophysics and Geodynamics, 151
\bibitem[Wietfeldt et al.(1996)]{wie96} Wietfeldt, R. et al. 1996, IEEE Transactions on Instrumentation and Measurement, 45(6), 923
\bibitem[Wilson, Roberts \& Davis(1996)]{wil96} Wilson, W., Roberts, P., Davis, E. 1996, in Proceedings of the 4th APT Workshop, ed E. King, 16
\end{thebibliography}
\end{document}